\documentclass[preprint,10pt]{sigplanconf}

\usepackage{amsmath}
\usepackage[LY1]{fontenc}
\usepackage[scaled=0.9]{helvet}
\usepackage{textcomp}
\makeatletter 
\input{ly1enc.def}
\makeatother
\DeclareTextSymbolDefault{\textplusminus}{LY1}
\usepackage{xspace}
\usepackage{multicol}
\usepackage{fancyvrb}
\usepackage{url}
\usepackage{graphicx}
\usepackage{color}
\usepackage{alltt}
\usepackage{floatflt}
\usepackage{subfigure}
\usepackage{microtype}

\newbox\subfigbox	
\makeatletter
\newenvironment{subfigenv}
{\def\caption##1{\gdef\subcapsave{\relax##1}}%
\let\reallabel\label
\def\label##1{\gdef\sublabsave{##1}}%
\let\subcapsave\@empty
\begin{lrbox}{\subfigbox}}
{\end{lrbox}
\subfigure[\subcapsave]{\usebox{\subfigbox}\reallabel{\sublabsave}}}
\makeatother

\definecolor{deletedtext}{rgb}{1,0.5,0.5}
\DeclareTextFontCommand{\textdel}{\color{deletedtext}}

\definecolor{annoteblue}{rgb}{0,0,1}
\DeclareTextFontCommand{\textnote}{\color{annoteblue}}

\newif\ifanonymous
\anonymousfalse

\newcommand{\name}[2]{\ifanonymous #1\else #2\fi} 

\newif\ifnotes
\notestrue

\newif\ifalt
\alttrue

\ifnotes
\newcommand{\cas}[1]{\textnote{\textbf{[CAS: #1]}}}
\newcommand{\meo}[1]{\textnote{\textbf{[MEO: #1]}}}
\newcommand{\jm}[1]{\textnote{\textbf{[JM: #1]}}}
\newcommand{\amc}[1]{\textnote{\textbf{[AMC: #1]}}}
\newcommand{\sap}[1]{\textnote{\textbf{[SAP: #1]}}}
\newcommand{\sab}[1]{\textnote{\textbf{[SAB: #1]}}}
\newcommand{\mjd}[1]{\textnote{\textbf{[JD: #1]}}}
\newcommand{\group}[1]{\textnote{\textbf{[GROUP: #1]}}}
\else
\newcommand{\cas}[1]{}
\newcommand{\meo}[1]{}
\newcommand{\jm}[1]{}
\newcommand{\amc}[1]{}
\newcommand{\sap}[1]{}
\newcommand{\sab}[1]{}
\newcommand{\mjd}[1]{}
\newcommand{\group}[1]{}
\fi

\let\cTX\texttt
\let\cTY\texttt
\let\cRW\texttt
\let\keyterm\emph
\let\prim\texttt 

\newcommand{\atomic}{\cRW{atomic}\xspace}
\newcommand{\spawn}{\cRW{spawn}\xspace}
\newcommand{\yield}{\cRW{yield}\xspace}

\newcommand{\blockuntil}{\cRW{blockUntil}\xspace}
\newcommand{\retry}{\cRW{retry}\xspace}
\newcommand{\orelse}{\cRW{orElse}\xspace}
\newcommand{\unprotected}{\cRW{unprotected}\xspace}
\newcommand{\yieldUntil}{\cRW{yieldUntil}\xspace}
\newcommand{\yielduntil}{\cRW{yieldUntil}\xspace}

\newcommand{\synchronized}{\cRW{synchronized}\xspace}

\newcommand{\Pthreads}{\textrm{Pthreads}\xspace}
\newcommand{\Pth}{\textrm{Pth}\xspace}

\newcommand{\Tinystm}{\textrm{TinySTM}\xspace}
\newcommand{\TLtwo}{TL2\xspace}

\newcommand{\Java}{Java\xspace}
\newcommand{\Naive}{Na\"{\i}ve\xspace}
\newcommand{\naive}{na\"{\i}ve\xspace}

\newcommand{\OCM}{Observationally Cooperative Multithreading\xspace}

\begin{document}

\conferenceinfo{PPoPP '11}{February 12--16, San Antonio, TX.} 
\copyrightyear{2011} 
\makeatletter
\def \@formatyear {2011/4/8}
  \makeatother

\copyrightdata{[to be supplied]}


\title{Observationally Cooperative Multithreading\thanks
{This material is based upon work supported by the National Science
  Foundation under Grant No.
  \ifanonymous XXX-XXXXXXX\else CCF-0917345\fi. 
  Any opinions, findings, and conclusions or recommendations expressed in this material are those of the authors and do not necessarily reflect the views of the National Science Foundation.}}

\date{2011/04/08}
\copyrightyear{2011}
\ifanonymous
\authorinfo{[Author(s) Omitted --- Anonymized Submision]\\\ } 
          {[Institution(s) Omitted]} 
           {[Email address(es) Omitted]} 

\else
\authorinfo{Christopher A. Stone \and
            Melissa E. O'Neill \and 
            Sonja A. Bohr \and
            Adam M. Cozzette \\
            M. Joe DeBlasio \and
            Julia Matsieva \and
           Stuart A. Pernsteiner \and
            Ari D. Schumer}
          {Harvey Mudd College}
           {\{stone,oneill,sbohr,acozzette,mdeblasio,jmatsieva,spernsteiner,aschumer\}@cs.hmc.edu}
\fi

\maketitle

\begin{abstract}

\iftrue 
Despite widespread interest in multicore computing, concurrency
models in mainstream languages often lead to subtle, error-prone code.

\keyterm{\OCM{}} (OCM) is a new approach to 
shared-memory parallelism. Programmers
write code using the well-understood cooperative (i.e., nonpreemptive)
multithreading model for uniprocessors. OCM then allows threads to 
run in parallel, so long as results remain consistent with the cooperative model.

\else

\sab{This first paragraph is still teh suck, but less teh suck than before.}
Hindering the potential benefits of parallel computing, mainstream concurrency
control mechanisms can be prohibitively hard to reason about and debug. To
ameliorate this problem we have developed \keyterm{\OCM{}} (OCM)---a parallel
programming model in which programmers write code in the style of the
well-understood cooperative (i.e., nonpreemptive) multithreading model for
uniprocessors. The underlying OCM implementation then optimizes
execution---running threads in parallel when possible---but in such a way that
the results are guaranteed to be consistent with CM.
\fi

\iftrue
Programmers benefit because they can
reason largely sequentially. Remaining interthread interactions are far less
chaotic than in other models, permitting easier reasoning and
debugging. Programmers can also defer the
choice of concurrency-control mechanism
(e.g., locks or transactions) until \emph{after} they have written their programs, at which point they can compare concurrency-control strategies and choose the one that offers the best performance.
Implementers and researchers also benefit from the agnostic nature of OCM---it provides 
a level of abstraction to investigate, compare, and combine a variety of interesting concurrency-control
techniques.
\else

OCM allows programmers to reason about parallel code in a natural, sequential
 way. Also, since an execution under OCM corresponds to some execution under
 CM, the model can be 
exploited for debugging purposes. Furthermore, OCM allows programmers to write
code which is not bound to a particular concurrency-control mechanism (such as
locks or transactions). This means that not only a particular application can
be easily ported to use a different underlying mechanism, but that an
intelligent OCM system can dynamically switch between different mechanisms
during a single execution. Researchers of concurrency-control mechanisms can
also benefit from the implementation-agnostic nature of OCM because it provides
a level of abstraction for easily comparing a variety of synchronization
techniques.
\fi


\end{abstract}





\section{Introduction} 

Parallel programming is notoriously difficult; 
it is hard to predict all ways in which threads may
interact~\citep{Lu-et-al:08:lrnng-mstks:sigpl}. Synchronization code to manage these interactions can be complex and
error-prone. And when bugs inevitably arise, hard-to-reproduce race conditions
make debugging even more difficult than in sequential code.
Although there has been valuable progress in making parallel programming
more accessible, the models for parallelism in widespread use today are
still difficult for many programmers to use effectively~\citep{Rossbach-et-al:10:is-trnsctnl:ppopp}.

Inspired by traditional Cooperative Multithreading (CM) for
uniprocessors, where threads run one at a time and continue until
they explicitly yield control, we propose a new model for parallel
programming. \keyterm{\OCM} (OCM) offers
\begin{itemize} 
\item Simple semantics and syntax, taken from CM;
\item Parallel execution, taking advantage of modern hardware;
\item Implementation flexibility, allowing a variety of contention management methods (including transactional memory and lock inference);
\item Serializability, simplifying debugging and reasoning. 
\end{itemize}

OCM is not an implementation mechanism, but rather an abstraction for
programmers. The observable behavior of programs is consistent with
execution on a uniprocessor with cooperative multithreading, even if behind the
scenes threads are running simultaneously or preempting one another.

Designed to emphasize correctness over raw performance, OCM may not be
suitable for all multithreaded applications. But just as many systems
use garbage collection and runtime bounds checking
rather than manual memory management and unsafe array accesses, we feel that
there is a place for systems like OCM that provide an easier and
safer path into parallel programming. And, as with garbage collection and
bounds checking, there is wide scope for interesting research and design work to
mitigate runtime overhead in OCM systems.
\goodbreak


In this paper, we define OCM and examine some of the issues involved in implementing and using
it in practice. In particular, we 

\begin{itemize} 
\item Introduce and define the core OCM model
(Section~\ref{sect:ocm});
\item Share our experience in building several prototype
 implementations of the OCM model (Section~\ref{sect:implementations});
\item Demonstrate that different concurrency-control mechanisms
can be compared by running the same algorithm on different
OCM implementations
(Section~\ref{sect:performance});
\item Describe how the OCM model can enable simple,
reproducible debugging of parallel programs (Section~\ref{sect:debugging}).
\end{itemize}

Readers should be aware that there are also some things that they will
\emph{not} find in this paper. In particular, because OCM is
a new model for shared-memory parallelism, there is no preexisting
storehouse of OCM programs, and thus no large benchmark suite
to run (see Section~\ref{sect:conclusion}). Our goal is not to show
how \emph{well} we have implemented OCM (although, anecdotally, we do
think it does perform well), but to show there are awkward parallel
programs that are more easily expressed in OCM, to demonstrate OCM's
potential, and perhaps encourage you to download an implementation and
experiment, or implement OCM with your own concurrency-control
mechanism.\footnote{Downloads are available at \url{www.ocm-model.org}.}



\ifalt

\section{Background}\label{sect:issues-with-existing-models}




Programmers working with mainstream languages already choose from a number of
multithreading models. The choice matters because it
can significantly affect whether programmers produce code free of race
conditions and deadlock~\cite{Rossbach-et-al:10:is-trnsctnl:ppopp}.

As a prelude to introducing our own model (in Section~\ref{sect:ocm}),
we review the strengths and weaknesses of some
well-known schemes for multithreading. We use the
familiar example of two
threads transferring funds within an array of bank
accounts, either with each thread doing exactly one transfer,

\noindent
\begin{minipage}{\columnwidth}
\vspace{1ex}
{\columnseprule=0.4pt\small

\begin{multicols}{2}
\begin{Verbatim}
 // move $5
 acct[x] = acct[x] - 5;
 acct[y] = acct[y] + 5;
\end{Verbatim}
\columnbreak
\begin{Verbatim}
// move $10
acct[i] = acct[i] - 10;
acct[j] = acct[j] + 10;
\end{Verbatim}
\end{multicols}
}
\vspace{0ex}
\end{minipage}
\noindent 
or with each thread looping to perform as many transfers as the
relevant accounts
permit:

\noindent
\begin{minipage}{\columnwidth}
\vspace{1ex}
{\columnseprule=0.4pt \small

\begin{multicols}{2}
 \begin{Verbatim}
while (acct[x] >= 5) {
  // move $5
  acct[x] = acct[x] - 5;
  acct[y] = acct[y] + 5;
}
\end{Verbatim}
\columnbreak
\begin{Verbatim}
while (acct[i] >= 10) {
  // move $10
  acct[i] = acct[i] - 10;
  acct[j] = acct[j] + 10;
}
\end{Verbatim}
\end{multicols}
}
\vspace{-1.5 ex}
\end{minipage}

\noindent We would like to ensure that money is neither created nor destroyed,
and that the loops cannot cause accounts to become overdrawn.


\subsection{Serial Computation with Cooperative Multithreading}
\label{sect:CM}

Cooperative Multithreading (CM) is a well-known model for writing
\emph{uniprocessor} multithreaded programs. In the CM model, exactly
one thread runs at a time, and control switches from one thread to
another only when a thread either terminates or uses the provided
\yield{} statement. Programmers place \yield{} statements at
specific points in the code where it is safe to yield
control---either to propagate changes to shared data between
threads, or just to be a ``good citizen'' and let other threads
execute.

Under CM, the two code examples above (nonlooping and looping) work
perfectly well as written.  Neither thread invokes \yield, so in both
cases first one thread will run to completion and then the other, with
no interleaving of computations.

In the looping case, if \cTX{x} happened to be equal to \cTX{i}, one looping thread would
transfer out most or all of the money, leaving little or nothing for
the other thread to do. We can provide an opportunity for interleaved
iterations by adding explicit \yield{} statements:

\noindent
\begin{minipage}{\columnwidth}
\vspace{1ex}
{\columnseprule=0.4pt \small

\begin{multicols}{2}
\begin{Verbatim}
 while (acct[x] >= 5) {
   // move $5
   acct[x] = acct[x] - 5;
   acct[y] = acct[y] + 5;
   yield;
 }
\end{Verbatim}
\columnbreak
\begin{Verbatim}
while (acct[i] >= 10) {
  // move $10
  acct[i] = acct[i] - 10;
  acct[j] = acct[j] + 10;
  yield;
}
\end{Verbatim}
\end{multicols}
}
\vspace{-1ex}
\end{minipage}

\noindent Because the \yield{}s are at the end of
the loops, individual iterations execute without interruption. 
In the first thread, for example, there is no possibility of 
\texttt{acct[x]} changing between the comparison and the assignments.

\paragraph{Critique} 

There are certainly instances where relying on cooperation
is inappropriate. Desktop operating systems such as Windows
and MacOS formerly employed CM but have long since adopted
preemptive scheduling, preventing one uncooperative process from
hanging the entire system. But within a single program, 
a buggy thread failing to \yield is no worse than an accidental
infinite loop in sequential code.

As a programming model for multithreaded applications, CM has
some very attractive properties. Most notably, the text of the program
specifies exactly where threads may be interrupted. Although CM programs may be
nondeterministic, the ways in which nondeterminism can arise
are relatively restricted.  Further, between \yield{}s we can
reason about code purely sequentially: until a thread \yield{}s,
it will never see the environment being changed by other threads, nor
will its changes be visible to other threads.

\enlargethispage{1em}
The composability properties of CM are also quite strong.  Two
nonyielding operations invoked in sequence automatically form a larger
nonyielding combination.  Programmers can also programatically control
whether code \yield{}s or not (e.g.,
having subroutines test a boolean flag to decide whether they should \yield).

We are not the first to see the CM model as a generally desirable
model for
programmers~\cite{von-Behren-et-al:03:evnts-ar:hotos,Gustafsson:05:thrds-wtht:queue}.
In recent years, it has been suggested most often to
combine the efficiency and simplicity advantages of sequential
event-driven code with a more natural programming
model~\cite{von-Behren-et-al:03:cprcc-sclbl:sosp,Fischer-et-al:07:tsks-lngg:pepm}.
There is still debate about the relative overheads of
events and cooperative threads, but there is no doubt that CM
can provide an attractive and natural model for systems
programming.

The obvious deficiency of CM as a model for multithreaded
computation is that it executes only one thread at a time. Our goal for OCM,
which we introduce in Section~\ref{sect:ocm}, is to retain the
many advantages of CM but allow multicore implementations. Before
turning to OCM, however, we contrast CM's relative simplicity
with some other common approaches to
parallel shared-memory computation.
\goodbreak

\subsection{Parallel Computation with Explicit Locking}

Parallel computation with explicit locking is one of the oldest parallel models
\cite{Dijkstra:68:cprtng-sqntl:plnasi}, and is still widely seen today
(\Pthreads, \Java threads, etc.). In this model, multiple
threads may execute simultaneously and control may switch between them at
unpredictable times. Locks and condition variables, monitors, semaphores, and
other similar explicit mechanisms provide concurrency control for shared data.  In this section, we will focus on locks.

In our banking examples, observe that money was transferred between accounts specified by variables (\cTX{x}, \cTX{y}, \cTX{i}, \cTX{j}), whose values we may not know until runtime.
One thread cannot assume that the other thread
will not try to access or modify one of the same accounts.
Therefore, code that works under CM has 
race conditions under the traditional preemptive model.  Consider our nonlooping example with \cTX{x} and \cTX{i} referring to the same account---our hope is that a total of \$15 is removed from that account, but for some executions it might decrease by only \$5 or \$10 (e.g., if the
second thread executes in its entirety between the first thread's read
of \cTX{acct[x]} and its write to \cTX{acct[x]}).  To
fix this issue, we can add locks to ensure that only one thread at a
time can
update a particular account. If we na\"{\i}vely add locks, our example becomes:\\
\noindent
\begin{minipage}{\columnwidth}
\vspace{1ex}
{\columnseprule=0.4pt \small

\begin{multicols}{2}
  \begin{Verbatim}
 lock(acct[x]);
 lock(acct[y]);
   // move $5
   acct[x] = acct[x] - 5;
   acct[y] = acct[y] + 5;
 unlock(acct[y]);
 unlock(acct[x]);
\end{Verbatim}
\columnbreak
\begin{Verbatim}
lock(acct[i]);
lock(acct[j]);
  // move $10
  acct[i] = acct[i] - 10;
  acct[j] = acct[j] + 10;
unlock(acct[j]);
unlock(acct[i]);
\end{Verbatim}
\end{multicols}
}
\end{minipage}\vspace{1ex}

\enlargethispage{1em}
\noindent But this code is incorrect, being
prone to deadlock if \cTX{x == j} and \cTX{y~== i}. There
are well-known techniques to prevent such deadlocks
(e.g., acquiring locks in a global total order); but doing so is 
not always simple. In this case, we cannot know whether to lock
\cTX{acct[x]} or \cTX{acct[y]} first until we know
the values of \cTX{x} and \cTX{y} at runtime; the problem
becomes worse when there are more locks or the locks are not
all acquired in one place.


\label{sect:background-locking-issues}

\paragraph{Critique}

Explicit locking comes with an interesting trade-off. A coarse-grained
locking scheme that holds a few locks for a long time (in the limit,
``one big lock for everything'') may operate correctly but offer
mediocre performance. Finer-grained schemes where
locks are held for as little time as possible may offer good
performance but be harder to reason about.


Because holding locks limits concurrency, this model often tempts the
programmer to write code with race conditions and then add
as few locks as possible, held for the shortest time
possible.  This choice can easily lead to bugs, as it is difficult to
mentally model all possible ways that complex lock-based code might execute.

For example, in the case of the looping account transfers, a programmer might 
acquire and release locks inside the loop (making each individual
transfer atomic), and even ensure that locks are acquired
in a good order, but not realize this strategy 
permits account contents to change between the loop
test and the body of the loop (permitting overdrawn accounts).   Correct
lock-based code even for these simple loops requires great care.





Finally, lock-based code does not compose well. If \cTX{foo} and
\cTX{bar} are each atomic because they acquire and release (possibly
different) locks, there may be no obvious way to combine both 
into a single atomic sequence that acquires the
correct locks in the correct order~\cite{Harris-et-al:05:cmpsbl-mmry:ppopp}.
 
\subsection{Parallel Computation with Atomic Blocks}
\label{sect:atomic}

An increasingly popular alternative to locks is the \atomic 
block~\citep{Lomet:77:prcss-strctrng:pacmc}.  As the name suggests,
\atomic blocks guarantee that the enclosed code appears to execute
atomically, even as multiple threads execute in parallel.
We can easily express the nonlooping example as

\noindent
\begin{minipage}{\columnwidth}
\vspace{1ex}
{\columnseprule=0.4pt \small

\begin{multicols}{2}
 \begin{Verbatim}
atomic {
  // move $5
  acct[x] = acct[x] - 5;
  acct[y] = acct[y] + 5;
}
 \end{Verbatim}
\columnbreak
\begin{Verbatim}
atomic {
  // move $10
  acct[i] = acct[i] - 10;
  acct[j] = acct[j] + 10;
}
\end{Verbatim}
\end{multicols}
}
\vspace{-1ex}
\end{minipage}


\paragraph{Critique}

Atomic blocks are usually easier to use than explicit locks, as we do not have
to worry about which locks to acquire or in what order, or which
condition variable to wait on.
With \atomic blocks, there is nothing else to specify---the
implementation infers all data dependencies.

Atomic blocks permit multiple underlying implementations. Two popular schemes 
are to infer and acquire the required locks or to use software transactional 
memory (discussed as a parallel model in its own right in 
Section~\ref{sect:stmintro}).

Another advantage of \atomic blocks over explicit locks is that
\atomic blocks 
are easy to compose. We can
combine individually atomic actions \cTX{foo}  and \cTX{bar} into
an atomic sequence simply by placing them inside an \atomic block.
But this approach requires explicit action by the programmer, and
one might easily forget that a sequence of atomic operations 
is not itself automatically atomic.

\enlargethispage*{1em}

Further, just as programmers working with locks may feel pressured to
hold locks for as short a time as possible, programmers working with
\atomic blocks may feel pressured to keep their \atomic sections
short, keeping as much code as possible in the ``unprotected'' outside
area, again increasing the chances that the programmer will introduce
errors.

\label{sect:background-atomic-probs}

Finally, the \atomic-block construct is not always as elegant
in practice as it may first appear.  Adapting the looping example to
use \atomic blocks, permitting interleaving but keeping the tests and assignments atomic,
demonstrates how much less intuitive and straightforward this model
can be: 
\vspace{0.5ex}

\noindent
\begin{minipage}{\columnwidth}
\vspace{1ex}
{\columnseprule=0.4pt \small

\begin{multicols}{2}
 \begin{Verbatim}
 bool loop1;
 do {
  atomic {
   loop1 = acct[x] >= 5;
   if (loop1) {
    // move $5
    acct[x] = acct[x] - 5;
    acct[y] = acct[y] + 5;
   }
  }
 } while (loop1);
\end{Verbatim}

\columnbreak

\begin{Verbatim}
bool loop2;
do {
 atomic {
  loop2 = acct[i] >= 10;
  if (loop2) {
   // move $10
   acct[i] = acct[i] - 10;
   acct[j] = acct[j] + 10;
  }
 }
} while (loop2);
\end{Verbatim}
\end{multicols}
}
\vspace{-1ex}
\end{minipage}

\subsection{Parallel Computation with Software Transactional Memory}
\label{sect:stmintro}
\label{sect:problems-with-stm}

Although software transactional memory (STM) 
may be used as an implementation
technique for
\atomic blocks, it is often seen as a parallelism model in its own right,
not necessarily tied to \atomic blocks.

In this model, shared data is accessed inside \keyterm{transactions}
(analogous to database transactions). Once a transaction starts, reads and
writes of shared data operate atomically with respect to other transactions
(i.e., as if they are taking place in isolation). When a transaction ends,
the STM system will attempt to \keyterm{commit} the transaction, but such a
commit may \keyterm{fail} due to conflicting changes made by other concurrent
transactions. In this case, it is \keyterm{rolled back} (i.e., all its work is undone) and may be retried.

The STM model provides an extension to the concept of \atomic blocks, but one in
which the programmer is given more control over the (now mandatory)
implementation mechanism. Atomic blocks may be provided under an STM model,
equating the opening brace of an \atomic block with \prim{beginTransaction}
and the closing brace with \prim{endTransaction}.
Users of STM
libraries might also begin and end transactions directly; for example,
implementing the looping code as:

\noindent
\begin{minipage}{\columnwidth}
\vspace{1ex}
{\columnseprule=0.4pt \small

\begin{multicols}{2}
 \begin{Verbatim}
 beginTransaction(); 
 while (acct[x] >= 5) {
   // move $5
   acct[x] = acct[x] - 5;
   acct[y] = acct[y] + 5;
   endTransaction();
   beginTransaction();
 }
 endTransaction();
\end{Verbatim}
\columnbreak
\begin{Verbatim}
beginTransaction(); 
while (acct[i] >= 10) {
  // move $10
  acct[i] = acct[i] - 10;
  acct[j] = acct[j] + 10;
  endTransaction();
  beginTransaction();
}
endTransaction();
\end{Verbatim}
\end{multicols}
}
\vspace{-1.5 ex}
\end{minipage}


\paragraph{Critique}

The STM model allows other constructs beyond \atomic blocks.  For
example, Automatic Mutual Exclusion (see Section~\ref{sect:related-ame})
provides \unprotected blocks, in which the opening of the block corresponds
to \prim{endTransaction} and the end of the block to \prim{beginTransaction}.


Programmers may further be allowed or
encouraged to take advantage of other transaction-based facilities,
such as the ability to explicitly \retry (i.e., fail) transactions, or
chain transactions together with an \orelse construct that begins a
second transaction only if the first decides to \retry{}.

Encouraging programmers to construct their own novel and elaborate
synchronization schemes using \retry and \orelse may also create
composability issues.  It is also unclear that such schemes are any easier to reason
about than similarly elaborate uses of locks and condition variables
\citep{Rossbach-et-al:10:is-trnsctnl:ppopp}.

As with explicit locks and \atomic blocks, transactional schemes potentially
allow code outside a transaction to access shared data without any concurrency
control at all. In some schemes, there may be additional pressure to perform
operations (such as I/O) outside of transactions because they cannot be
rolled back. For instance, the following code could result in undesired
behavior:
{\small
\begin{Verbatim}
 atomic {
   print("Hello World");  // Do some I/O
   retry;                 // Undo that I/O
 }
\end{Verbatim}
}
\noindent
Therefore, an STM system might require that I/O take place outside of
transactions.

Although transactions are a powerful mechanism, as a parallel model,
they can, like locks and atomic blocks, lead to programs that are
intricate, subtle, and hard to reason about.

\subsection{Conclusions}

The parallel models we have discussed can be difficult to use due to a
number of factors. They all allow, encourage, or sometimes even require
programs to have portions that run outside the provided concurrency-control
mechanisms.  They each have subtleties that may trip up unwary or inexperienced
parallel programmers, and may encourage programmers to use the provided
concurrency-control mechanisms in intricate or fragile ways. 
\citet{Rossbach-et-al:10:is-trnsctnl:ppopp} found that undergraduate students
in a systems class had more difficulty understanding the transactional
concurrency model than they did the coarse-grained locking model,
but those who used locks produced programs with significantly more errors.

``Conscious
human thinking appears to us to be sequential, so that there is something
appealing about software that can be considered in a sequential 
way''~\citep{Skillicorn-Talia:98:mdls-lnggs:cmpsrv}.
The CM model provides significant opportunities for sequential reasoning in
an explicitly multithreaded program. It is, we argue, easier
to reason about, has fewer subtleties, and allows \emph{nothing} to run outside
of the provided concurrency-control mechanism. Unfortunately, the 
concurrency-control mechanism it provides is harsh indeed---no parallelism at all, only serial interleaving of threads at \yield points. We
desire a system that provides the benefits of CM \emph{and} the parallelism of
the other models. 


\else
\input{background.tex}
\fi


\section{Observationally Cooperative Multithreading}
\label{sect:ocm}

We therefore propose a model for parallel computation called Observationally
Cooperative Multithreading (OCM). It adopts the simple semantics of
cooperative multithreading (CM) discussed in
Section~\ref{sect:CM}. Unlike CM, the OCM model allows implementations that
take advantage of multiprocessor parallelism when possible.

As with CM, under the OCM model the programmer simply specifies locations in
their code where it is safe for a thread to yield control; the syntax for an
OCM program is the same as for a CM program. For example, the previous banking
example written for a CM system is also a correct example written in the style
of OCM:

\noindent
\begin{minipage}{\columnwidth}
\vspace{1ex}
{\columnseprule=0.4pt \small

\begin{multicols}{2}
\begin{Verbatim}
 while (acct[x] >= 5) {
   // move $5
   acct[x] = acct[x] - 5;
   acct[y] = acct[y] + 5;
   yield;
 }
\end{Verbatim}
\columnbreak
\begin{Verbatim}
while (acct[i] >= 10) {
  // move $10
  acct[i] = acct[i] - 10;
  acct[j] = acct[j] + 10;
  yield;
}
\end{Verbatim}
\end{multicols}
}
\vspace{-1ex}
\end{minipage}

But unlike CM, OCM is a model for \emph{parallel} computation. A system implementing
the OCM model is free to run programs in parallel, provided that the observable
behavior (final results, I/O, etc.) of a program is consistent with a possible
execution under some (nonpreemptive, uniprocessor) CM model. We call this
requirement \keyterm{CM serializability}, and it is the fundamental property of
OCM.

For the above code, the two loops can execute simultaneously if \cTX{x} and
\cTX{y} are disjoint from \cTX{i} and \cTX{j}, or be serialized
otherwise; either way produces results consistent with CM.

CM serializability also means that the semantics
of OCM is by definition that of CM; we can immediately
reuse existing formalizations of CM
semantics~\cite{Abadi-Plotkin:09:mdl-cprtv:popl},
and hence omit formal semantics here.

\subsection{A Parallel Perspective on the OCM Model}

From a parallel-execution perspective, code
between any two dynamically successive \yield{} statements executes atomically in OCM. 
Threads behave as if completely isolated from each other
except at \yield points. Thus, \yield statements should be placed
where a thread needs to publish its changes to the surrounding 
environment and/or to observe other threads' changes to that environment.

The details of how an OCM system runs code in parallel while retaining CM
serializability (and the concurrency-control mechanisms it uses to do so) are
implementation decisions, visible to users only insofar as they affect
performance. Like \atomic blocks, OCM may be implemented using a variety of
concurrency-control schemes, which can range from basic to elaborate. We
examine these options in detail in Section~\ref{sect:implementations}.

\subsection{Advantages of the OCM Model}

OCM offers programmers the same advantages as CM, particularly the ability to
reason about parallel threads in serial chunks punctuated by \yield statements.
Yet it also avoids CM's main disadvantage: support for only
uniprocessor execution. In addition, OCM benefits from being agnostic
about the underlying mechanism.

This agnosticism means, first of all, that a programmer using the OCM
model can avoid making a premature commitment to any particular
concurrency-control scheme.  This flexibility can be particularly
useful if a programmer is not sure beforehand whether their
application will work best with optimistic concurrency control (e.g.,
STM) or pessimistic concurrency-control (e.g., locks).  If a program
is written in OCM, it can be easily ported to OCM systems that provide
the same interface but very different underlying implementations.
Thus a programmer can test out which mechanism suits the application
best.  The same program written with explicit locks or transactions
would make this comparison much more difficult.

Also, external code written outside the OCM model can generally be
used in conjunction with an OCM program, so long as the OCM system
treats it conservatively, which means treating external code as
possibly \yield{}ing and/or having appropriate conflicts.
For example, we can handle
unbuffered I/O operations by going to either extreme: reduce parallelism
by serializing I/O access to ``the world'' (i.e., if another thread is doing I/O, we must wait for it to yield before we can do I/O),
or maximize parallelism by treating
all I/O as  \yield{}ing before and after. In the latter case,
the proviso ``\texttt{fscanf} will \yield{}''
can be compared to a typical STM restriction that ``calls to
\texttt{fscanf} may not appear in an \atomic block.''

\subsection{Beyond \yield{}}
\label{subsect:idioms} 
OCM is consistent with many traditional concurrency primitives,
including mutexes,
condition variables, and barriers as in the GNU Pth library for
CM~\cite{Engelschall:06:GnuPth:manual}. If we worry that relying solely on shared-memory
concurrency might not scale well to huge numbers of processors, then
an OCM implementation can provide channels and
primitives for threads to do synchronous or asynchronous
message-passing.

One approach is to implement these primitives 
directly using \yield{} and shared data.  For example,
{
\small
\begin{alltt}
  do
    yield;
  while (!\ensuremath{p})
\end{alltt}
}
\noindent
begins a conditional critical
region~\citep{hoare:72:theory-of-parallel:book,harris+:03:lightweight-trs:oopsla};
any following code executes atomically with the test $p$, once $p$
is true.  

This idiom is very powerful, and all our prototype
OCM implementations provide \yielduntil{}\texttt{(\ensuremath{p})} as a built-in 
operator. We can use it, for example, to implement a simple barrier:
{
\vbox{
\small
\begin{alltt}
 void barrier() \{
   ++count;
   \yielduntil ( count == NUM_THREADS );
 \}
\end{alltt}
}
}
\noindent
The OCM code for a more robust and reusable barrier is a little longer, but
easily achievable.
\goodbreak

A large number of interesting kinds of coordination and
synchronization mechanisms can be explained in terms of
\yield. Although we can implement them in this fashion, OCM
implementors also have the option of writing more sophisticated and
efficient native implementations. Programmers using OCM can largely
ignore the difference; it would remain valid to imagine a thread at a barrier repeatedly \yield{}s until everyone arrives.  An STM-based implementation,
however, might implement \yieldUntil as
{ 
\small
\begin{alltt}
  endTransaction();
  beginTransaction();
  while (!\ensuremath{p}) retry;
\end{alltt}
}
\goodbreak
\noindent because a sophisticated STM might 
record the shared variables used to evaluate the predicate
\textit{p} and, if $p$ is side-effect free, delay \retry{}ing the
transaction until one of those variables
has changed~\citep{Harris-et-al:05:cmpsbl-mmry:ppopp}.


\subsection{Trade-Offs of the OCM Model}

CM serializability imposes a ``concurrency control everywhere and always''
requirement that may impact performance for some programs. However, this
trade-off is deliberate and, we believe, potentially worthwhile if
it can improve simplicity and correctness.

In addition, by providing an abstraction that hides the details of the
underlying concurrency-control mechanism, OCM makes invisible any unique
features that would violate CM serializability.  Anything not easily realizable
under CM (such as rolling back execution to an earlier point) will not be
exposed. Thus, even if the underlying implementation of OCM is transactional,
features such as \retry or \orelse are hidden. This simplification may limit the
level of control the programmer has over their program's execution, but
OCM implementations (and support libraries) may use these features behind the scenes to provide efficient of
CM-compatible concurrency primitives.

Finally, like CM, STM, \atomic blocks, and explicit locking, OCM does not
guarantee determinism or the absence of race conditions; multithreaded code remains harder than unthreaded code.

\subsection{Fairness and Uncooperative Threads}
\label{sect:yield-fairness}

Most implementations of CM provide a \yield-fairness
guarantee: any thread that is suspended because it invokes
\yield will eventually be allowed to resume. This fairness guarantee
holds only in the absence of \keyterm{uncooperative} threads, threads
that neither \yield{} nor terminate. When an uncooperative thread
exists, it either will be scheduled to run (in which case it will
monopolize the CPU forever, unfairly halting all other threads), or
the uncooperative thread itself is being unfairly avoided.

In this paper, when referring to CM, we implicitly assume
\yield fairness in the absence of uncooperative threads.  (It would be
difficult to program under the assumption that an unfair scheduler
might arbitrarily refuse to resume a pariticular thread.) An uncooperative thread in a CM
program are almost certainly a programmer error, so permitting
unfairness in this case seems reasonable. Unlike some
languages where programmer errors mean that all bets are off
(``undefined behavior''), there remains a well-defined semantics
for programs with uncooperative threads,
though with restricted possibilities for interleaving.
CM serializability thus requires that in the absence of uncooperative
threads, all \yield{}ing OCM threads will eventually resume.

In a CM system, if an uncooperative thread does exist and begins (observably) executing,
it will prevent all other threads from executing. This situation could occur in an OCM
implementation as well. But if the uncooperative thread has no observable effects (i.e., no I/O, no shared-data changes, etc.) it is possible that OCM
could run it forever on one processor while other processors execute the
remaining threads.  As \yield fairness is not guaranteed in the
presence of uncooperative threads, we still remain consistent with a CM
implementation; specifically, one that is unfair to the uncooperative
thread. 

\section{Implementations} 
\label{sect:implementations}

Because OCM is implementation agnostic, a variety of techniques can be used to
develop a valid OCM system.  We have developed several different implementations, which are all available for download at \url{http://ocm-model.org/}. In creating these implementations, we show that a variety of implementation strategies for OCM are feasible. Doing so also allows us to compare the tradeoffs of these different implementation strategies. In this section, we discuss some of the salient issues that arise from implementing OCM; in Section~\ref{sect:performance} we will see how these implementations compare in practice.


\subsection{\Naive Implementations}
\label{sect:naive}

Possibly the simplest implementation of the OCM model is traditional
uniprocessor CM. Although CM does not exploit multiple cores, it has
value as a baseline implementation. We would hope that an OCM
implementation that exploits multiple cores would quickly outperform CM.
But sometimes CM may actually be the best OCM implementation to use (e.g., 
for programs with massive thread contention, or on a uniprocessor machine).

\Naive parallel implementations are also possible. One such scheme is to use a
preemptive implementation with a single global lock to protect all
shared data. In this case, \yield could be implemented as
\prim{releaseGlobalLock} followed by \prim{acquireGlobalLock}. Like CM, only
one thread would run at a time, but different threads might execute on different
cores.

Interestingly, however, the global-lock scheme can be optimized by delaying 
\prim{acquireGlobalLock} until shared data is about to be accessed for
the first time since \yield{}ing.  Likewise, if the system can determine that a
\yield{} is imminent and that no more accesses to shared data will occur before
the next \yield{} is reached, it can perform the \prim{releaseGlobalLock}
action ahead of the actual \yield. We call these two lock optimizations
\keyterm{lazy acquire} and \keyterm{eager release}.  Both optimizations
preserve CM serializability, yet allow thread executions to overlap in
parallel.

Lazy acquire is fairly trivial to implement, but eager release
seems to require static analysis.  To avoid or enhance static analysis,
an OCM system can allow the programmer to make assertions about the
behavior of their code.  Thus, a programmer may assert that the thread will yield
within a specific amount of time 
and that the program will not use any more shared data
until that \yield.  Both assertions can be checked and enforced at run
time, and in a
global-lock OCM implementation, they may provide enough information for the
system to release the lock early.

Using a single global lock is hardly cutting-edge concurrency control, but,
like CM, it provides a baseline against which more sophisticated
concurrency-control schemes may be compared. Furthermore, it successfully
provides some parallelism at low implementation cost.

\subsection{Nontrivial Lock-Based Implementations}
\label{sect:lock-impl}

The OCM model can be implemented with far more sophisticated lock-based
implementations than one global lock.
Much of the work on lock inference for atomic blocks is directly applicable to
the problem of executing OCM threads (see Section~\ref{sect:related-lockinf}). In contrast, in this
section we outline a simple scheme based on per-object locks to show that this
approach is feasible as an OCM implementation.

As before, the OCM source program does not refer to locks.  But if 
the thread is accessing a shared data, the OCM system must (
on the thread's behalf) acquire the proper locks before the
thread's access occurs, and release them at the following \yield.
This use of locks guarantees that one thread
can never modify an object that is in use by another thread, so a running
thread will never observe outside changes to a shared variable between two
\yield statements.


As with most situations that use mutual-exclusion locks, a lock-based OCM
implementation must have some mechanism to handle or prevent deadlock.  One
solution is to impose a global total order for acquiring the locks,
although doing so requires that the OCM system know in advance which variables
a thread might use between each pair of \yield statements.  Conservative predictions
can be obtained via static analysis, augmented by runtime state information.


Like our earlier \naive global-lock--based implementation, our system can make
use of programmer-specified assertions to optimize its behavior.  For example,
if the programmer asserts that a section of the program will access only a
specific range of indices in a shared array, the OCM implementation might
choose to use fine-grained locking to allow threads that require access to
other parts of the same array to run concurrently.  In addition, such
assertions can be checked at runtime to detect errors in the program.
Similarly, any lock-based scheme can exercise lazy lock acquisition and eager
lock release. The only added caveat is that all locks needed between yields
must be acquired before any may be released, two-phase locking
\cite{Eswaran-et-al:76:ntns-cnsstncy:cacm}.

\subsubsection{Dynamic-Language Implementation}
\label{lua-proof-of-concept}

We have developed a proof-of-concept lock-based OCM implementation as an
extension to the Lua scripting language~\cite{Ierusalimschy-et-al:06:l-5:book}.
This extension is a dynamic library loaded by the Lua interpreter, so it cannot
perform static analysis to obtain the information needed for correct locking.
 
Access to shared data is therefore mediated solely through ``proxy objects'' 
obtained through the OCM library---threads are otherwise completely separate. 
Because the system knows that a thread can only access shared data through
proxies, and the system knows which threads are holding which proxies,
the OCM scheduler can acquire all necessary locks for a thread before
it can run.

The above approach is conservative; just because a thread has expressed
interest in a shared value by acquiring its proxy doesn't mean that the
thread will necessarily access it before the next \yield. Lazy acquire
can be implemented by waiting to acquire locks until the appropriate proxy is
accessed (subject to lock-ordering constraints). Eager release needs
the program to tell the system about proxies that will not be accessed
before the next \yield, using informational function calls (much like \cTX{assert} is used in C and C++). Run-time checks ensure that these assertions are accurate.

I/O performed by threads must either be to separate files, or also mediated through the OCM library.

This implementation shows that it is workable to create a relatively light-weight
OCM extension to an existing language without making significant changes to the language core. But there is a trade-off: this OCM implementation is more syntactically awkward than some of the others we will describe, because declaring and accessing shared data requires function calls. Runtime confirmation of the programmer's assertions imposes further overhead.


\subsubsection{Compiled-Language Implementation}
\label{source-to-source-translator}

We have also implemented lock-based OCM in the form of a source-to-source
translator for a simple C-like language, a restricted form of
C/C++ with the addition of \yield and \spawn statements.  The translator analyzes
the original source code to conservatively determine which variables may be accessed in the future following each \yield statement---those are the variables that \yield needs to lock.  This information is then used to insert calls to
locking and unlocking functions using \Pthreads in the necessary locations. 
Any \spawn or \yield statements are also replaced with calls to library functions.
The translator has options to enable lazy locking as well as simple static analysis for eager unlocking.

There is already much work on static analysis to automatically perform \emph{lock inference} for atomic blocks \cite{Ringenburg-Grossman:05:atmcml-frstclss:icfp,
  Hindman-Grossman:06:atmcty-v:mspc,
  Emmi-et-al:07:lck-allctn:popl,Cherem-et-al:08:infrrng-lcks:pldi,KhilanGudka:10:hll-wrld:tr}, but analyzing a program that \yield{}s is not quite the same as analyzing one with more traditional \atomic blocks. First, \atomic blocks have a statically scoped endpoint, whereas the location of the next \yield may be dynamically determined. Second, \atomic blocks have code outside of the \atomic block, whereas there is no equivalent code in OCM. Thus, while our proof-of-concept is no doubt far less sophisticated than existing lock-inference schemes, there is value in showing that it can be done for the OCM model.


\subsection{STM-based Implementations}
\label{stm-implementations}

The OCM model also permits implementations based on
software transactional memory.  In such an implementation, all reads
and writes of shared data are routed through an STM system.  Each \yield
statement ends the current transaction and begins a new one, 
so that changes made by the current thread become visible to
others.  

Most STM libraries are designed for use with transactions that begin and end in
the same lexical scope, so they require modification for use in OCM
implementations.  For example, OCM may require that a transaction begin in one
function and end in another, as shown in the following example:

\enlargethispage{0ex}
{\small\vspace{0.5ex}%
\newcommand{\backupabit}{\vspace{-1ex}}
\begin{alltt}
  void subroutine() \{
    yield;
    \vdots 
  \}
\end{alltt}
\begin{alltt}
  void caller() \{
    subroutine();
    \vdots
    yield;
  \}
\end{alltt}
\vspace{0ex}
}

If the transaction ending at the \yield at the end of \cTX{caller} 
cannot commit, it must roll back to the \yield inside
\cTX{subroutine}, which has since returned and been removed from the stack.  Thus,
the STM system needs to be able to ``unreturn'' from functions
when a transaction aborts and retries. Unreturning is not conceptually difficult; it simply requires some state saving so that the stack can be restored if a transaction fails~\cite{Smaragdakis-et-al:07:trnsctns-isltn:oopsla}.

Using STM as an implementation technique also presents problems with
I/O and other operations with side effects, which cannot be reversed
if a transaction needs to be rolled back.  One method commonly used in STM
systems is to require that a thread wishing to perform I/O obtains a
special lock which guarantees that its transaction always succeeds. Another
possibility is to have the OCM system buffer output until a transaction
completes successfully and print it before starting the next transaction.
This technique is effective, but it cannot be applied to input.  A final method
is to force an implicit \yield before and after I/O operations.  This option allows the
I/O to be done without rollback.

As we noted in Section~\ref{sect:problems-with-stm}, it is unclear what an STM
implementation should do in the event of nested transactions. STM-based OCM
implementations avoid this problem because every statement
takes place in exactly one transaction.


\subsubsection{Dynamic-Language Implementation}
\label{lua-proof-of-concept-stm}

As when investigating lock-based implementations, we began with a
proof-of-concept modification to Lua.  In this case, we implemented the OCM
system by requiring the Lua interpreter to use the 
\Tinystm \citep{Felber-et-al:08:dynmc-prfrmnc:ppopp} library when
accessing global variables.  We also modified it to support \Pthreads and ``unreturning''
from Lua functions by tracking changes to the interpreter stack so
that they can be rolled back if needed.

Our implementation experience here reveals that it is possible to adapt an existing scripting language to mediate all of its data accesses through an STM system, although the changes required can be quite invasive. But the effort comes with a positive pay-off---to language users, access to shared data is simple and natural.


\subsubsection{Compiled-Language Implementation}
\label{sect:cplusplus-library}

We have also created an STM-based OCM implementation as a C++ library using \Pthreads.
This library allows the programmer to indicate that certain global variables are
shared, which causes all accesses to those variables to be routed through
either the \TLtwo or \Tinystm systems.  Our library includes implementations
of \yield and 
\spawn, and supports ``unreturning'' from
functions by transparently saving portions of the stack.

Our library approach requires no changes to the underlying language, relying instead on C++ language features (overloading, templates, etc.) to make access to shared data feel natural. As with our dynamic-language OCM system, most of the hard work for concurrency control is done by the STM system, but unlike that system, much of the implementation work is mere shimming. In principle, a transactional approach could also benefit from statically derived information about program behavior, but like more traditional STM systems, our implementation does not perform any static analysis.

Extending an STM library is a quick way to implement a parallel OCM system, can provide one that is highly usable in practice (certainly no more difficult than using an STM library directly), and allows complex programs to be expressed naturally and run in parallel.



\begin{figure*}
\mbox{}\hfill
\begin{subfigenv}
\begin{minipage}{0.45\linewidth}\small
\begin{verbatim}
philosopher(int i):
  for iter in (1..ITERS):
    think();
    yield;
      
    eat(fork[i], fork[(i+1) % N])
    yield;    
\end{verbatim}
\vfill\mbox{\ }


\caption{Philosophers never observe each other holding forks.}
\end{minipage}
\label{subfig:phils-invisible}
\end{subfigenv}
\hfill\hfill
\begin{subfigenv}
\begin{minipage}{0.45\linewidth}\small
\begin{verbatim}
philosopher(int i):
   for iter in (1..ITERS):
      think();
      yieldUntil (isFree[i] && isFree[(i+1) % N]);
      
      isFree[i]         = false;   // take left fork
      isFree[(i+1) % N] = false;   // take right fork
      yield;
      eat();
      yield;
      isFree[i]         = true;    // put down forks
      isFree[(i+1) % N] = true;
      yield;    
\end{verbatim}\vspace{-0.5ex}
\caption{Philosophers can observe each other holding forks.}
\end{minipage}
\label{subfig:phils-visible}
\end{subfigenv}
\hfill\mbox{}
\caption{Solutions to the Dining Philosophers Problem}
\label{fig:phils}
\end{figure*}

\subsection{Other Implementation Techniques}

The only requirement OCM places on its implementation is that it conform to CM
serializability. In Sections~\ref{sect:lock-impl}
and~\ref{stm-implementations}, we examined lock-based and STM-based schemes,
but other schemes are possible. An OCM system could, for example, use a hybrid
of locks and transactions, defaulting to STM-based concurrency control while
having the option to fall back on a lock-based implementation or even CM in the
case of high contention for shared data.


Because an OCM implementation can combine statically and dynamically gathered
information, possibly augmented with programmer assertions about the behavior
of their code, other interesting concurrency-control options may be possible.
For example, lock-free techniques such as atomic processor instructions or
sequential locking could be used to improve performance. Consider a program that
contains a shared
variable \cTX{counter} that is always used to initialize a local variable as follows:
\begin{Verbatim}
 yield;
 int timestamp = ++counter;
 yield;
\end{Verbatim}
In this case, it may be safe to avoid protecting \cTX{counter} with
locks or transactions and use a processor instruction for atomic increment.


\begin{figure*}
\vspace{-1.75em}
\mbox{}\hfill
\subfigure[Algorithm from Figure~\ref{subfig:phils-invisible}, speedup relative to CM.\label{subfig:perf-invisible}]{\includegraphics[scale=0.42]{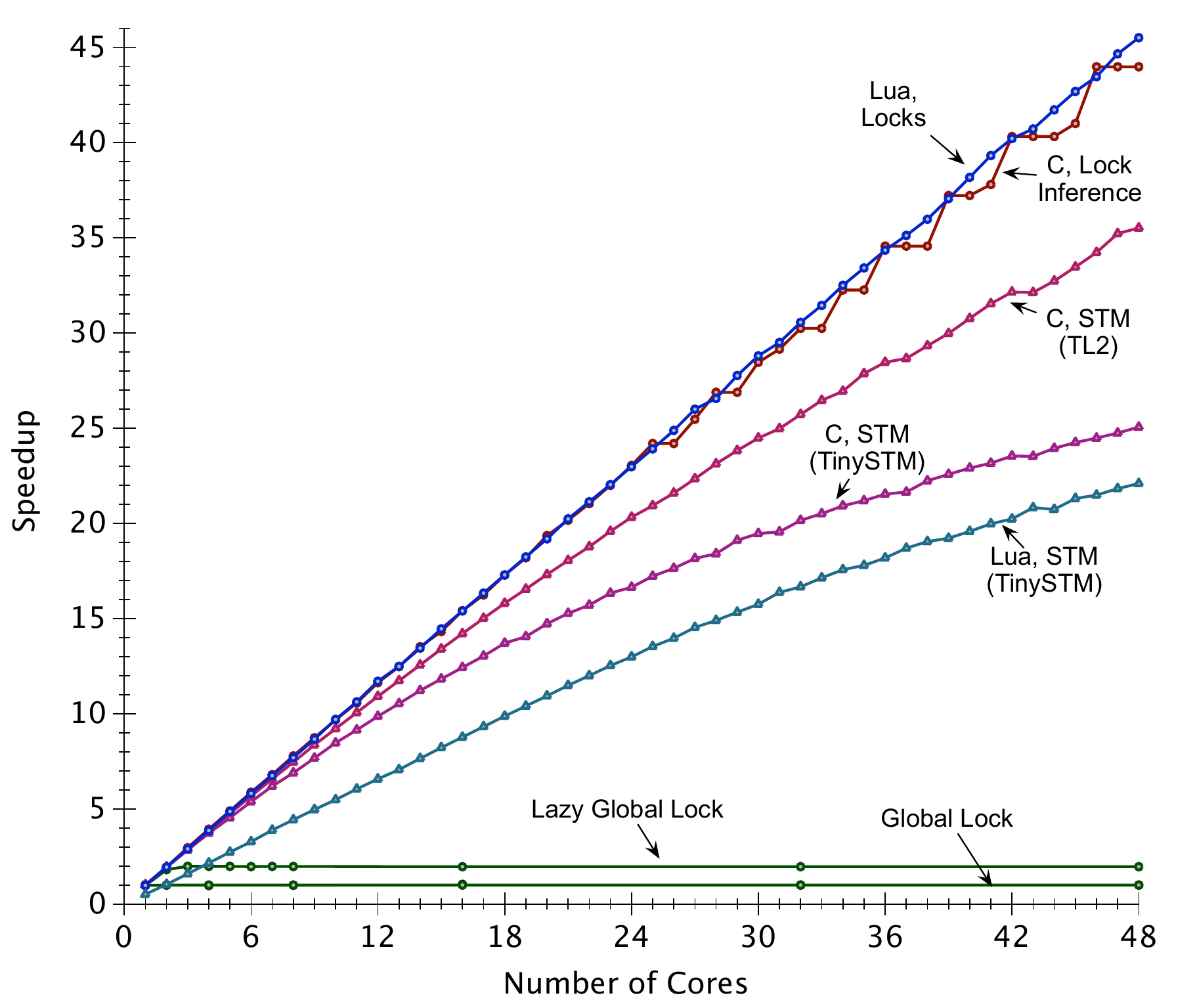}}
\hfill\hfill
\subfigure[Algorithm from Figure~\ref{subfig:phils-visible}, speedup relative to CM.\label{subfig:perf-visible}] {\includegraphics[scale=0.42]{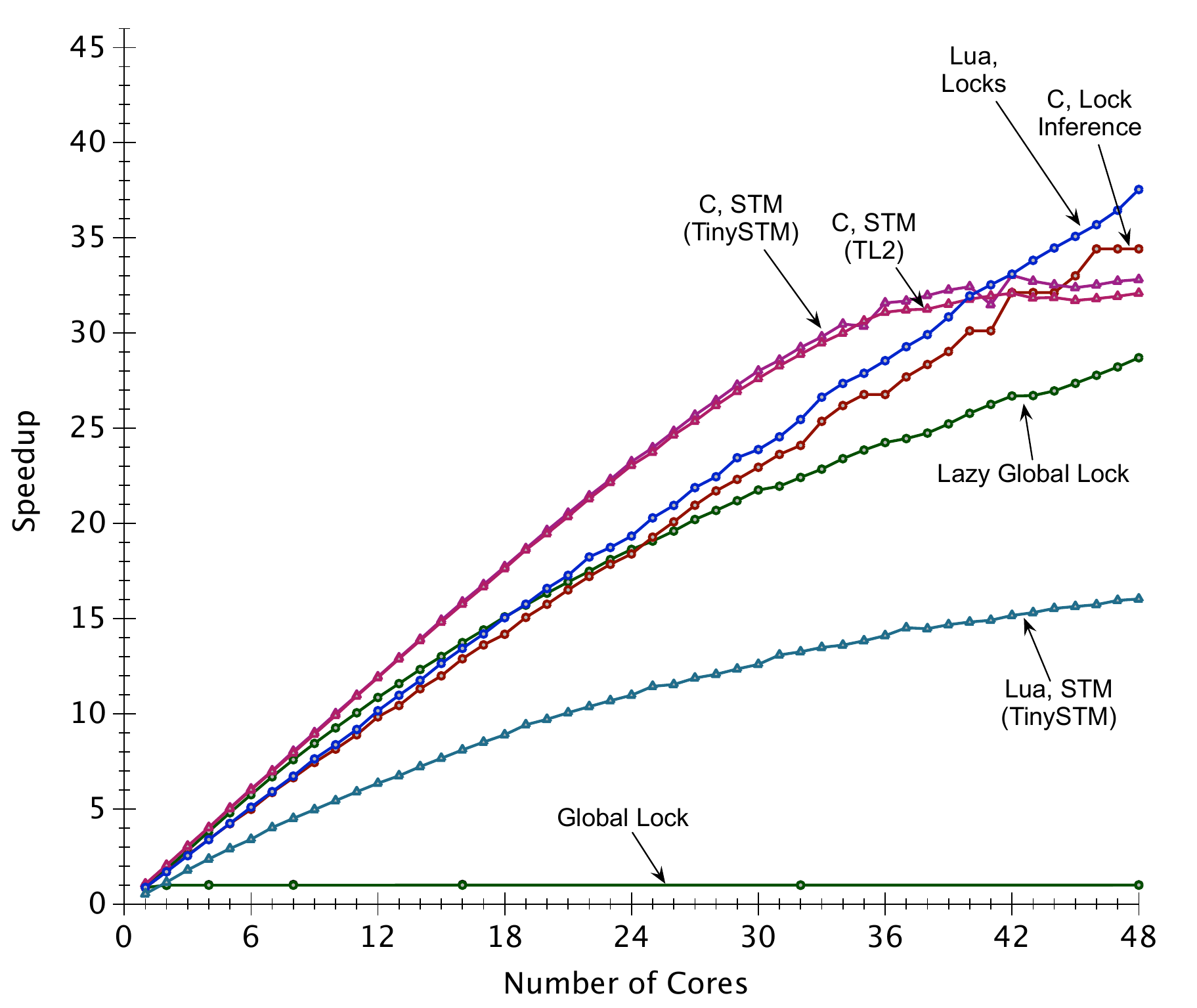}}
\hfill\mbox{}
\caption{Performance of Our Dining Philosophers Example.}
\label{fig:graphs}
\end{figure*}




\section{Comparing Implementation Strategies}
\label{sect:performance}

In the previous section, we showed that the OCM model is
implementable; in this section, we show that it is possible to compare
different concurrency-control techniques underlying the same OCM
program.

We therefore turn to that classic problem in concurrency, \emph{dining philosophers}%
\footnote{Tanenbaum \cite{Tanenbaum:*:modern-os:book} writes that
  ``everyone inventing a new synchronization primitive has tried to
  demonstrate how wonderful the new primitive is by showing how
  elegantly it solves the dining philosophers problem,'' making our
  choice almost \emph{de rigueur}.
}~\cite{Dijkstra:68:cprtng-sqntl:plnasi,Dijkstra:1965:tw-strvtn:tr}.
First suggested 
in 1965%
, the problem is still 
studied to this day \cite{Danturi-et-al:09:stblzng-phlis:taas,Downey:08:lttl-sems:book}.
In the problem, \cTX{N} philosophers are arranged around a table,
alternating between \emph{thinking} and \emph{eating}. Eating is complicated, as
each philosopher requires two utensils to eat (e.g., forks or chopsticks), but each utensil must be shared by two neighboring philosophers. A solution to the problem should avoid deadlock, livelock, and other forms of starvation.
 
Although dining philosophers might seem overfamiliar, it is an
easy-to-understand problem whose solutions are often intricate, and
one where some widely seen and taught solutions have unexpected
subtleties.  Gingras \cite{Gingras:90:dnng-phlsphrs:sigcse} observes
that Tanenbaum's semaphore-based solution
\cite{Tanenbaum:*:modern-os:book} is not strictly starvation-free, but
the behavior of Gingras's starvation-avoiding solution is also
somewhat nonobvious~\cite{Yue:91:dnng-phlsphrs:sigcse}. Despite its
apparent simplicity, the problem
provides interesting insight into both the OCM model and its implementations.

For our tests, we used two OCM-based solutions to the problem.
Figure~\ref{subfig:phils-invisible} shows an almost trivial solution.
In a CM implementation, only one philosopher would
ever eat at a time, with no interference between neighbors; CM
serializability guarantees an indistinguishible result under OCM.
In this solution, the philosophers
are literally oblivious to each other: while philosophers are using
their forks, they do not \yield, so at every \yield point all
philosophers
see all forks on the table. It
is up to the OCM system to find and exploit parallelism, and ensure 
(because of \yield-fairness) that every
philosopher makes progress without any philosophers starving.  The code
itself is also interesting because it is virtually identical to code that is
usually presented as an unworkable race-prone solution attempt \cite{Tanenbaum:*:modern-os:book}; the only
difference is the added \yield{}s and the requirement that it run under the OCM
model.

Figure~\ref{subfig:phils-visible} shows a more involved solution where philosophers can
see their neighbors holding forks, and hence must explicitly wait for
their own forks to become
available. (Eating itself remains a private affair).
The added \yield{}s give the OCM implementations more latitude
for thread interleaving. 

This second solution is constructed
to be parallel to the first, but is actually prone to the same issue as
Tanenbaum's semaphore based solution
\cite{Gingras:90:dnng-phlsphrs:sigcse,Yue:91:dnng-phlsphrs:sigcse}---a thread
can be starved if its neighbors happen to alternate their eating in a way that
always overlaps. We might argue that the simplicity of our code makes the flaw
easier to notice, but also note that the perpetual pathological interleaving necessary to starve a philosopher is
unlikely in practice. There are a variety of ways to ensure there is no chance
of starvation (specifically, avoid situations where we eat
again while our neighbor is hungry), but the simplest solution is to use the
code in Figure~\ref{subfig:phils-invisible}.

To allow the problem to scale as we add processor cores, our tests use 199
philosophers rather than the more typical five. In addition, our timings are for
1000 iterations (per philosopher)---a reduction from the usual infinite number
of iterations.

\goodbreak

Figure~\ref{fig:graphs} shows the results of running the two algorithm variants on a 48-core machine.%
\footnote{Specifically, a SuperMicro H8QGi+-F--based system, with four Opteron
6168 processors running at 1.9 GHz, and 64 GB of RAM running Linux (Ubuntu
10.04). Each processor MCM has two dies, each with six cores. (The effects of
the six-core boundaries are visible in some of the graphs.)}
All versions use the same delay loops to simulate eating and
thinking, calibrated to take about 1 ms \textplusminus{} 20\%{}
pseudorandom variation, chosen as a point where contention effects start to become visible enough to make the graphs ``interesting.''

We show speedup graphs for each algorithm, where speedup is measured compared to a
corresponding pure CM implementation---for our Lua-based code, it is an implementation using Lua's coroutine facility, and for our C-code implementations, the CM system is a thin wrapper around GNU \Pth~\cite{Engelschall:06:GnuPth:manual}.

In Figure~\ref{subfig:perf-invisible}, we see that both STM and lock-inference
schemes are finding the potential parallelism in this
problem. The performance of our lock-based OCM implementations is
nearly identical to that of well-known solutions
\cite{Tanenbaum:*:modern-os:book,Gingras:90:dnng-phlsphrs:sigcse} (not
shown on the graph).  Interestingly,
as we scale to multiple processors there is little difference between
C and Lua lock implementations. Our C-code transactional implementation can use either
TL2 \citep{Dice-et-al:06:trnsctnl-lckng:disc}  or TinySTM
\citep{Felber-et-al:08:dynmc-prfrmnc:ppopp}
as its STM back-end; for this program, TL2 seems to perform better
with the rather long transactions that arise from this algorithm. Our
STM-based Lua implementation is slower, presumably because considerably more
state is tracked in transactions for the interpreter than for C. Finally, as
you might expect, the \naive implementations offer little speedup, but
interestingly, both do run faster than their CM counterparts on a multiprocessor---the global lazy
lock offers a 1.8\textmultiply{} speedup on two cores, and about 2\textmultiply{} on more than two cores (staying essentially flat beyond three cores).

In Figure~\ref{subfig:perf-visible}, we see the performance for the
second variant of the problem. The lock-based implementations again
perform well, but are outperformed by both variants of our C-based
STM implementation (whose performance is so similar that we only show
one line on the graph) until we reach about 38 cores; beyond that point,
STM performance does not scale as well. The loss of scalability is largely due to
our use of a simple implementation of \yieldUntil---variants that use \retry
(not shown on the graph) scale much better, but don't perform as well in absolute time.

The global-lazy-lock implementation also
does well in this case. This good performance is largely due to the contrived nature of the task---because the code already explicitly does its own fork arbitration, \cTX{eat()} does not actually \emph{use} the forks in any way, and thus no shared data accesses occur during \cTX{eat()}, which in turn means that the global lock is never acquired during \cTX{eat()}.  Nevertheless, this version does show that sometimes na\"{\i}ve solutions perform well; for programs that do most of their computation independently and occasionally coordinate, the global-lazy-lock approach can work surprisingly well.

Readers should not suppose that this one simple program by itself
reveals anything particularly noteworthy about the relative strengths
of lock-based and transactional implementations, but what we have
shown is that we can compare their performance running essentially
\emph{the same program}. The comparison in this case reveals what we
might expect: which scheme wins out depends on a variety of factors,
including the number of cores and the structure of the program. What
is tantalizing about OCM is the potential it provides to adaptively choose the concurrency control scheme that works best; for example, for the second algorithm, we might choose a transactional approach for smaller numbers of cores and switch to locks for very large numbers of cores.

A secondary, largely anecdotal, result we can infer from our measured performance is that there exist programs for which our OCM implementations scale well. Although we do not devote space to it here (creating a full-blown benchmark-suite for OCM is a paper in itself), our results thus far with regard to scalability seem encouraging.

Readers interested in examining a slightly more realistic program written using OCM can consult the appendix to this paper.



\section{Debugging and Performance Profiling} 
\label{sect:debugging}

Although OCM dramatically reduces the potential for race conditions
and deadlock compared to, say, explicit locking, it does not
eliminate them.
The following code uses shared variables \cTX{a} and \cTX{b} to
(atypically) simulate explicit locks in
a way that could lead to deadlock, assuming both \cTX{a}
and \cTX{b} are initially \cTX{0}:
{\columnseprule=0.4pt
\iffalse
\begin{multicols}{2}
\noindent\emph{Thread A}
\begin{verbatim}
 do {
   yield;
 } while (a > 0);
 ++a;
 do {
   yield;
 } while (b > 0);
 ++b;
 yield;
 --a;
 --b;
\end{verbatim}
\columnbreak
\emph{Thread B}
\begin{verbatim}
 do {
   yield;
 } while (b > 0);
 ++b;
 do {
   yield;
 } while (a > 0);
 ++a;
 yield;
 --b;
 --a;
\end{verbatim}
\end{multicols}
\else
\small
\vspace{-1ex}
\begin{multicols}{2}
\begin{verbatim}
 yieldUntil (a == 0);
 a = 1;
 yieldUntil (b == 0);
 b = 1;
\end{verbatim}
\columnbreak
\begin{verbatim}
 yieldUntil (b == 0);
 b = 1;
 yieldUntil (a == 0);
 a = 1;
\end{verbatim}
\end{multicols}
\vspace{-2ex}
\fi
}
\noindent 
It is possible to write buggy multithreaded code in the OCM
model (\yield fairness does not guarantee that other operations like
\yieldUntil will succeed); debugging
and repeatability for parallel programs is a longstanding problem~\cite{Ronsse-DeBosschere:99:RecPlay:tocs,Bergan:10:coredet:asplos,Zyulkyrov:10:DebugAtomicTM:ppopp}.

Fortunately, reproducing bugs is far easier 
in OCM than in many other models due to reduced
opportunities for race conditions.
Further, in OCM every execution of a program has at least one corresponding
execution under CM. If an OCM system wishes to allow reproducible
debugging, it simply has to record a corresponding serial execution
for that program. With that \emph{serialization trace}, it is possible
to rerun the program serially following that trace and thereby
reproduce the exact sequence of interleavings that trigger the bug. 
We have implemented a proof of concept in our C++ STM-based
OCM implementation (described in Section~\ref{sect:cplusplus-library}).

For the code above, a human-readable example trace of a failed
execution might read as follows:
{\small
\begin{alltt}
 A->B (at A's `yieldUntil (a == 0);')
 B->A (at B's `yieldUntil (b == 0);')
 A->B (at A's `yieldUntil (b == 0);')
 B->A (at B's `yieldUntil (a == 0);')
 {\rmfamily\itshape... deadlock ...}
\end{alltt}
\vspace{-1ex}
}

From this trace it is fairly straightforward for programmers to work out what
has gone wrong, or for them to simply rerun a failed execution to better
understand what happened.

OCM implementations may run threads in
parallel, so how can they record a serial execution order that
corresponds to their parallel execution?  One scheme is to use a global timestamp service and have each thread record timestamps to create a trace. In the case of a lock-based implementation, the timestamp should be recorded after all locks have been acquired and before any lock has been released. In the case of a transactional implementation, the timestamp should be read during the transaction, and recorded when the transaction is successfully committed.
 
In fact, no system-level
support is necessary. We can achieve tracing just by adding one line
of code after every \yield to record the trace in a large shared array
using a shared index
{\small
\begin{verbatim}
 yield;
 trace[index++] = (thread_id, context_info);
\end{verbatim}
}
\noindent but for some OCM implementations, adding this extra code may restrict observable execution orders.
For this reason, a low-overhead system-level implementation is preferable.

Recording thread-serialization traces has more potential uses than
just supporting debugging. For example, knowing how often one thread runs
compared to others can also reveal and explain performance issues with
the code, such as when the underlying concurrency-control system fails to
achieve the desired level of concurrency.


\section{Conclusions and Future Work} 
\label{sect:conclusion}

OCM is a promising solution for shared-memory program
development. It retains many of the benefits of currently existing 
concurrency-control systems, while mitigating the complexity of using these systems. It
allows the programmer to concentrate more on the logic of the program and less
on the subtle mechanics of concurrency control.

As we have demonstrated with our lock-based and STM-based systems, the OCM
model can be implemented with various underlying concurrency-control systems. In
this way, an application can be written according to the OCM model and use
whichever implementation is best suited for it. It may even be worthwhile to refactor 
existing multithreaded programs to use OCM in order to make future development
 or debugging easier.

We would like to see OCM broadly adopted. As we have shown, it is often straightforward to
use OCM as a front end for a variety of concurrency-control mechanisms, and so we
hope that others will follow our lead and show how their concurrency-control schemes
can be used to execute programs written for OCM. We also hope that educators see
the value in using OCM as a ``kinder gentler'' form of multicore parallelism, even if they
later introduce other, more challenging, models such as explicit locks or transactions.
In fact, OCM can be a springboard for exploring these other techniques; synchronization
primitives are easy to write in OCM (e.g., \cTX{semWait($i$)} is \cTX{yieldUntil($i$ > 0); \textminus\textminus$i$}, ~and \cTX{semSignal($i$)} is \cTX{++$i$}), and discussions of efficient OCM implementations naturally lead
to topics like transactions. We hope
that our available implementations and further examples of OCM in use (which include solutions
to a number of other classic and not-so-classic problems \cite{Downey:08:lttl-sems:book})
will provide a good starting point for these efforts.

We are continuing our OCM implementation work, and we plan to take even
better advantage of modern research into concurrency control (see
Section~\ref{sect:ocm-overhead}). We have, of course, already run somewhat
larger examples than dining philosophers (with promising results), but
there is still much to learn about scaling efficiently OCM to large systems,
including which language extensions
and debugging and profiling tools prove most valuable.

In addition, OCM needs a suite of benchmark programs that can be used
to assess the performance of different concurrency control techniques
and of the OCM approach as a whole. Unfortunately, existing benchmark
suites are targeted at  prior schemes for parallelism, and although it
is possible to recreate other schemes within OCM (e.g., by rolling
your own semaphores, locks, or condition variables), doing so misses
the OCM's point of allowing simpler solutions.
Thus, new benchmarks must be created from scratch.




\section{Related Work}
\label{sect:related}

As a parallel model, OCM intersects with a significant portion of prior work on
parallelism and concurrency. There is a vast literature describing parallel
models, concurrency-control mechanisms, debugging techniques, and so forth that
could be compared to OCM, but we cannot hope to do them all justice here.
Thus, we must restrict our discussion to those techniques that we feel are
of most interest because they parallel, influence, or counterpoint OCM in a
particularly significant way. 

\subsection{Other Models of Concurrency}
 \label{sect:related-ame}
 There are, of course, many other models for parallel programming
 besides those discussed in
 Section~\ref{sect:issues-with-existing-models}, including
 monitors~\cite{Hoare:74:monitors:cacm} and Java \synchronized
 methods~\cite{Gosling:05:JLS:book}, communicating sequential
 processes~\cite{Hoare:85:CSP:book}, Threading Building
 Blocks~\cite{Reinders:07:TBB:book}, and
 OpenMP~\cite{OpenMP:08:OpenMPAPI3.0:spec}, to name just a few.  
 We cannot compare each in detail here, but to the extent that they provide
 particular scheduling policies or ways to create new threads
 (e.g., parallel \cRW{for} loops), they may be transferrable to an OCM
 context. However, three further models deserve specific comparison with
 OCM.

\paragraph{Automatic Mutual Exclusion}
 AME~\cite{Isard-Birrell:07:atmtc-mtl:hotos,Abadi-et-al:08:smntcs-trnsctnl:popl}
 is a variant of software transactional memory.  Rather than starting
 with unsynchronized code and marking particular blocks as \atomic,
 AME makes the safer assumption that all code should
 be executed atomically unless specifically marked
 \unprotected. Consequently, atomic code is dynamically delimited by
 the execution of \unprotected blocks.  An AME system could 
 easily be used to implement OCM (\yield corresponds to an empty
 \unprotected block~\cite{Abadi-et-al:08:smntcs-trnsctnl:popl}), 
 and AME has already engendered work on the denotational semantics of
 uniprocessor cooperative multithreading~\cite{Abadi-Plotkin:09:mdl-cprtv:popl}.

 While we have found the AME work inspiring, there are two
 ways in which OCM intentionally differs from AME.  Both follow
 naturally from CM serializability, but the differences
 make OCM both simpler and safer.

 First, AME exposes the underlying STM implementation. Its \blockuntil operator
 permits users to roll back and retry in the middle of an atomic
 transaction, allowing code that is impossible without run-time 
 tracking and undoing side-effects. In contrast, OCM has multiple
 implementations, including lock-based schemes that are more
 efficient for some programs.

 Second, OCM has no escape hatch for ``lower-level'' memory operations
 outside of OCM's concurrency-control system. Non-empty \unprotected
 blocks can improve performance, but they can also be a
 source of bugs and semantic
 surprises~\cite{Shpeisman-et-al:07:enfrcng-isltn:pldi}.


\paragraph{Transactions with Isolation and Concurrency}
Compared with traditional STM systems, the main feature of
TIC~\cite{Smaragdakis-et-al:07:trnsctns-isltn:oopsla} is
its ability to ``punctuate'' atomic transactions.  

The \texttt{\cRW{Wait}(\ensuremath{p})} statement checks whether $p$
is true; if not, it commits the current transaction, and waits
for $p$ to become true before continuing in a new transaction.
Transactions can begin and end in different functions; like
STM-based OCM (and like AME), TIC implementations must capture enough
of a run-time continuation to undo function-returns.

A motivating example for TIC is a transaction that waits on a barrier;
although the number of waiting threads should be incremented before we
wait for remaining threads to arrive, inside a transaction this
increment would not be visible to any other thread until the transaction
ends (after the barrier).  The TIC solution is to increment the count and
\cRW{Wait}; since \texttt{\cRW{Wait}(\ensuremath{p})} corresponds
exactly to 
\texttt{\cRW{while}\ (!\ensuremath{p})\ \yield;} in OCM,
OCM provides similar functionality.

TIC has other features not in OCM.  It does static checking to ensure
that calls to methods that might \cRW{Wait}\ are marked as such. (A
similar approach might be desirable for OCM methods that might
\yield.)  TIC also lets a program check at run time whether these
calls actually \cRW{Wait}, and take corrective action if necessary.
TIC is otherwise based on \atomic blocks, and its treatment
of nested transactions is slightly subtle. OCM has \yield but
not \atomic, and has no possibility of nesting.

\paragraph{Cilk}
Cilk~\cite{Blumofe-et-al:95:clk-effcnt:ppopp,Frigo-et-al:98:implmnttn-clk5:pldi98}
 is a parallel extension of C.
Although Cilk largely relies on the programmer to prevent interference
between threads, it is interesting to compare Cilk's \keyterm{serial
  elision} property with CM serializability. 

Serial elision
guarantees that Cilk keywords can be omitted---replacing 
\cRW{spawn}ed function calls with ordinary function calls,
and removing all \cRW{sync} barriers---to obtain a legal C program with
the same semantics as the Cilk version running on one processor.
The Cilk version running on a multiprocessor may produce
additional behaviors (due to race conditions and other
nondeterminism).

The CM serializability guarantee goes in the opposite direction:
the behaviors of OCM running on a multiprocessor cannot exceed
the behaviors possible in theory from a uniprocessor CM implementation.
\goodbreak

\subsection{Methods for Reducing OCM Overhead}
 \label{sect:ocm-overhead}
 \label{sect:related-lockinf}

There are many ways to implement OCM. For example, OCM can use a pessimistic lock-based approach. 
\keyterm{Lock Inference} is a method for the system to infer the correct locking actions
automatically \cite{Ringenburg-Grossman:05:atmcml-frstclss:icfp,
  Hindman-Grossman:06:atmcty-v:mspc,
  Emmi-et-al:07:lck-allctn:popl,Cherem-et-al:08:infrrng-lcks:pldi,KhilanGudka:10:hll-wrld:tr}. 
Research advances in lock inference can be directly applied to
lock-based OCM implementations.

 \label{sect:related-stm}
 There is also significant ongoing research into \keyterm{Software Transactional
  Memory}~\cite{Shavit-Touitou:95:sftwr-trnsctnl:podc,Larus-Rajwar:07:trnsctnl-mmry:slca,Spear:10:lghtwght-rbst:spaa}
(and even hardware support for transactional memory~\cite{Harris-et-al:05:cmpsbl-mmry:ppopp,Larus-Rajwar:07:trnsctnl-mmry:slca}). STM
 has even been adopted by some
newer languages such as Sun's
Fortress~\cite{Allen-et-al:08:frtrss-lngg:tr}.
Switching an STM-based OCM implementation from one STM library
to another is not
difficult, as long as the STM interfaces are reasonably similar. As a
result, an STM-based OCM implementation
can choose the STM implementation with the best performance
characteristics.

In general, implementation advances in other models of concurrency 
should permit improved implementations of OCM.

%

\acks We would like
to thank \name{AAA}{Bartholomew Broad}, \name{BBB}{Kwang Ketcham},
\name{CCC}{Samuel Just}, \name{DDD}{Alejandro Lopez-Lago}, and
\name{EEE}{Joshua Peraza} for
their work prototyping early OCM implementations,
and \name{EEE}{Christine Alvarado}, 
\name{FFF}{Claire Connelly}, and \name{GGG}{Robert Keller} for helpful
comments on this paper.


\bibliographystyle{abbrvnat}


{\softraggedright
\bibliography{ocm}
}
\appendix

\vfill
\section*{Appendix:  A Larger Sample Program, Ants}
The following code excerpt demonstrates runnable code in the C++ STM
implementation of OCM.  The code, inspired by Rich Hickey's ant colony simulation
for Clojure but significantly simplified, simulates ants wandering a grid looking for
food. Two ants must not occupy
the same square; this is ensured by not \yield{}ing between the
time an ant determines a square is free and it moves.
Similarly, the printing thread shows a consistent
view of the grid (e.g., not catching any ants in the middle of their move)
because the \cTX{print} method does not \yield.

The \cTX{board} variable is a global variable
of class \cTY{Grid} class, not shown.  It
contains a
\Verb|SharedArray<int> grid;| plus straightforward
helper functions such as
{\small
\begin{Verbatim}
char Grid::get(size_t x, size_t y) {
    if ( inRange(x,y) ) return grid[x + y * WIDTH];
    return '0'
}
\end{Verbatim}
}

\vspace*{2em}

\subsection*{Sample Output (showing one instant in time)}

\vspace{1ex}
{\fontsize{7.75}{8}\selectfont
\begin{Verbatim}
food count is 6
+ + + + + + + + + + + + + + + + + + + + + + + + + + + + + + 
+ . . . . . A . . . . . . . . . . . . . . . . . . . . . . + 
+ . . . . . . . . . . . . . . . . . . . . . . . . . . . . + 
+ . . . . . . . . . . . . . . . . . . F . . . . . . . . . + 
+ . . . . . . . . . . . . . . . . . . . . . . . . . . . . + 
+ . . . A . . . . . . . . . . . . . . . . . . . . . . . . + 
+ . . . . . . . . . . . . . . . . . . . . . . . . . . . . + 
+ . A . . . . . . . . . . . . . . . . F . . . . . . . . . + 
+ A . . . . . . . . . . . . . . . . . . . . . . . A . . . + 
+ . . . . . . . . . . . . . . . . A . . . . . . . . . . . + 
+ . A . . . . . . . . . . . A . . . . . . . A . A . . . . + 
+ . . . . . . . . . . . . A . . . . . . . . . . . . . . . + 
+ . F . . . . A . . . . . . . . . . . . . . . . . . . . . + 
+ . . . . . . . . . . . . . . . . . . . . . . . . . . . . + 
+ . . . . A . . . . . . . . . . . . . . . . . . . . . . . + 
+ . . . . . . . . . . . . . . . . . . . . . . . . . . . . + 
+ . . . . . . . . . . . . . . . . . . . . . . . . . . . . + 
+ . . . . . . . . . . . A . . . . . . . . . . . . . . . . + 
+ . . . . . . . . . . . . . . . . . . A . . . . . . . . . + 
+ . . . . . . . . . . . . . . . . . . . . . . . . . . . . + 
+ . . . . . . . . . . . . . . . . . . . . . . . . . . . . + 
+ . A . . . . . . . . . . . . . . . . . . . . . . . . . . + 
+ . . . . . . . . . . . . . . . A . . . . . . . . . . . . + 
+ . . . . . . . A . . . . . . . . . . F . . . . . A . . . + 
+ . . . . . . . F . . . . . . . . . . . . . . . . . . . . + 
+ . . . . . . . . . . . . . . . . . . . . . . . . . . . . + 
+ . . . . . . . . . . . . . A . . . . . . . . . . . . . . + 
+ A . . . . . . . . . . . . . . . . . . A . . F . . A . . + 
+ . . . A . . . . . . . . . . . . . . A . . . . . . . . . + 
+ + + + + + + + + + + + + + + + + + + + + + + + + + + + + + 
\end{Verbatim}

}

\pagebreak

\subsection*{Code Excerpts}


\small

\newcommand{\Yield}{{\bfseries\normalsize{}yield}}

\begin{alltt}
#include "ants.hpp"

void* ant (void* args) \{
  OCM_THREAD(&args);
  
  point * arg = (point *) args;
  int x = arg->x;
  int y = arg->y;
  ocm::ocmFree(arg); 
  
  int health = STARTING_HEALTH;
  
  while (1) \{
    \Yield;
    
    if ( board->getCount(FOOD) == 0 ) break;

    // Decide how to move (to food, or random)    
    point foodHere = findFood(x,y);
    int dx = foodHere.x - x;
    int dy = foodHere.y - y;
    
    if ( dx == 0 && dy == 0 ) \{
      dx = rand() % 3 - 1;
      dy = rand() % 3 - 1;
    \}
    
    if( board->traversable(x+dx,y+dy) ) \{
      // move
      board->set(x,y,EMPTY);
      x += dx;
      y += dy;
      if ( board->get(x,y) == FOOD ) 
         // eat
         health += FOOD_HEALTH;
      board->set(x,y,ANT);
    \} 
        
    \Yield;
    usleep(ANT_DELAY);
    
    health -= DECAY_RATE;
    if ( health <= 0 ) \{
      // Not enough food.
      board->set(x,y,DEAD_ANT);
      break;
    \}
  \}
  
  return NULL;
\}
\end{alltt}
\vfill\goodbreak

\begin{alltt}
point findFood(size_t x, size_t y) \{
  point looking = \{x,y\};
  
  // Check neighboring points for food
  for (size_t i = x-1; i <= x+1; ++i)
    for (size_t j = y-1; j <= y+1; ++j)
      if ( board->get(i, j) == FOOD ) \{
        looking.x = i;
        looking.y = j;
        return looking;
      \}
  
  return looking;
\}

// Repeatedly prints the current state of the grid.
void* printLoop(void * args) \{
  OCM_THREAD(&args);
  
  while(board->getCount(FOOD) > 0) \{
    board->print();
    \Yield;
    usleep(PRINT_DELAY);
    \Yield;
  \}

  return NULL;
\}


void addAnts(ocm::thread_t* &antThreads) \{
  point* p;
  int i = 0;
  while ( i < NUM_ANTS ) \{
    p = (point *) ocm::ocmMalloc(sizeof(point));
    p->x = rand() % GRID_SIZE;
    p->y = rand() % GRID_SIZE;
    
    if (board->get(p->x,p->y) == EMPTY) \{
      board->set(p->x,p->y,ANT);
      ocm::thread_create(&antThreads[i], NULL, ant, p);
      ++i;
    \} else 
      // Grid was occupied; try again
      ocm::ocmFree(p);
  \}
  
  return;
\} 
\end{alltt}
\vfill\goodbreak

\begin{alltt}
void addFood() \{
  size_t x,y;
  int foodCount = NUM_FOOD;
  
  while(foodCount > 0) \{
    x = rand() % GRID_SIZE;
    y = rand() % GRID_SIZE;
    if (board->get(x,y) == EMPTY) \{ 
      board->set(x,y,FOOD);
      --foodCount;
    \}
  \}    
  
  return;
\}

int main(int argc, const char * argv[]) \{
  OCM_START(NUM_ANTS+1);
  
  board = new Grid(GRID_SIZE,GRID_SIZE);
  addFood();
  board->print();
  
  ocm::thread_t* antThreads = new ocm::thread_t[NUM_ANTS];
  ocm::thread_t printThread;
  addAnts(antThreads);
  
  usleep(ANT_DELAY);
  
  ocm::thread_create(&printThread, NULL, printLoop, NULL);
  
  for (int i = 0; i < NUM_ANTS; ++i) \{
    ocm::thread_join(antThreads[i], NULL);
  \}
  
  ocm::thread_join(printThread,NULL);
  
  \Yield;
  
  board->print();
  
  delete[] antThreads;
  delete board;
  
  OCM_EXIT();
  
  return 0;
\}
\end{alltt}


\end{document}